\documentclass[a4paper]{jpconf}

\usepackage{amssymb}
\usepackage{mathptmx}       
\usepackage{helvet}         
\usepackage{courier}        
\usepackage{type1cm}        
\usepackage{makeidx}         
\usepackage{graphicx}        
\usepackage{multicol} 

\newcommand{\A}{{\mathcal{A}}}

\begin{document}
\input{epsf}
\title{Cosmic magnetic fields and dark energy in extended electromagnetism}

\author{Jose Beltr\'an Jim\'enez$^{\dagger, \ddagger}$ and Antonio L. Maroto$^\dagger$}

\address{$^\dagger$Departamento de F\'isica Te\'orica, Universidad Complutense 
de Madrid, 28040, Madrid, Spain\\ 
$^\ddagger$ Institute de Physique Th\'eorique, Universit\'e de Gen\`eve, 24 quai 
E. Ansermet, 1211 Gen\`eve 4, Switzerland}

\ead{Jose.Beltran@unige.ch, maroto@fis.ucm.es}

\begin{abstract}
We discuss an extended version of electromagnetism in which the usual gauge fixing term is promoted into a physical contribution that introduces a new scalar state in the theory. This new state can be generated from vacuum quantum fluctuations during an inflationary era and, on super-Hubble scales, gives rise to an effective cosmological constant. The value of such a cosmological constant coincides with the one inferred from observations as long as inflation took place at the electroweak scale. On the other hand, the new state also generates an effective electric charge density on sub-Hubble scales that produces both vorticity and magnetic fields with coherent lengths as large as the present Hubble horizon.

\end{abstract}

\vspace{-0.5cm}

\section{Introduction}

Out of the four known fundamental interactions in nature, two of them are particularly interesting in cosmological contexts due to their long range action, namely: gravity and electromagnetism. In fact, the behaviour of these two interactions at large scales is far from clear. On the gravitational sector we find the intriguing problem of the cosmic acceleration. Although a cosmological constant provides a 
simple and accurate description of it, from
a theoretical point of view it would be even more desirable to have
a fundamental explanation for the tiny value of such a  constant.
In this sense, several modifications of the gravitational interaction on 
cosmological scales have been proposed in the literature \cite{ModGrav}. Concerning the electromagnetic sector, the unknown origin of the $\mu$G magnetic
fields observed in galaxies and clusters \cite{Widrow} and, more remarkably,   
the very recent claim of detection of extra galactic magnetic fields
\cite{extragalactic} still lacks of a satisfactory explanation.

In this work we will consider 
the potential role of a modified electromagnetic theory 
in the dark energy problem \cite{EM1,EM2,EM3} and how this modified electromagnetism can generate magnetic fields at large scales.  Thus, we shall explore the interesting possibility of finding
a link between dark energy and the origin of cosmic magnetic field.

\section{Extended electromagnetism without the Lorenz condition}

In the covariant quantization of the electromagnetic field, the starting point is the modified Maxwell action including a gauge fixing term, that, in a curved spacetime, reads:
\begin{equation}
S=\int d^4x\sqrt{-g}\left[-\frac{1}{4}F_{\mu\nu}F^{\mu\nu}+\frac{\xi}{2}(\nabla_\mu A^\mu)^2+J_\mu A^\mu)^2\right]
\label{actionF}
\end{equation}
which leads to the modified Maxwell equations:
\begin{equation}
\nabla_\nu F^{\mu\nu}+\xi\nabla^\mu(\nabla_\nu A^\nu)=J^\mu.
\label{EMeqexp}
\end{equation}
Taking the divergence of this equation, we obtain:
\begin{equation}
\Box(\nabla_\nu A^\nu)=0
\label{minimal}
\end{equation}
The quantization in the Gupta-Bleuler formalism and in Minkowski spacetime proceeds as follows \cite{Itzykson}: One works with the four polarizations of $A_\mu$ and then imposes the weak Lorenz condition on the physical states $\vert\phi\rangle$ so that $(\partial_\mu A^\mu)^{(+)}\vert\phi\rangle=0$ to get rid of the unphysical polarizations. With this condition, the expected value of any physical observable only depends on the the transverse degrees of freedom, because the temporal and longitudinal polarizations contribute with opposite signs and, since the Lorenz condition imposes that there must be the same number of temporal and longitudinal photons in any physical state, they cancel each other. However, in an expanding universe, we see from (\ref{minimal}) that $\nabla_\nu A^\nu$ can be excited from vacuum quantum fluctuations because it behaves as a massless scalar field and, therefore, the Lorenz condition is violated. The reason for such a violation is that the expansion excites a different number of temporal and longitudinal photons so the aforementioned cancellation does not occur anymore \cite{EM1,EM2}.

In order to avoid the difficulties found when quantizing in the covariant formalism in an expanding universe, let us explore the possibility that the fundamental 
theory of electromagnetism is  given 
 by the modified action (\ref{actionF}), where we allow
the $\nabla_\mu A^\mu$ field to propagate. This theory still has the residual gauge symmetry $A_\mu\rightarrow A_\mu+\partial_\mu\theta$ provided $\Box\theta=0$.
 Thus, having removed one constraint, 
the theory  contains one additional degree of freedom and
the general solution for the modified equations
can be written as:
\begin{eqnarray}
\A_\mu=\A_\mu^{(1)}+\A_\mu^{ (2)}+\A_\mu^{(s)}+\partial_\mu \theta
\end{eqnarray}
where $\A_\mu^{(i)}$ with $i=1,2$ are the two transverse modes of
the massless photon, $\A_\mu^{(s)}$ is the new scalar state, which
is the mode that would have been eliminated if we had imposed the
Lorenz condition and, finally, $\partial_\mu \theta$ is a purely
residual gauge mode, which can be eliminated by means of a
residual gauge transformation in the asymptotically free regions, 
in a completely analogous way to the elimination of the $A_0$
component in the Coulomb quantization.  The fact that
Maxwell's electromagnetism could contain an additional scalar
mode decoupled from electromagnetic currents, but with 
 non-vanishing  gravitational interactions, was already noticed 
in a different context in \cite{Deser}. 

The evolution of the new mode is given by (\ref{minimal}), so that  
 on super-Hubble scales,
$\vert\nabla_\mu\A^{(s)\mu}_k\vert= const.$ which, as shown in
\cite{EM1}, implies that the field contributes as a cosmological
constant in (\ref{actionF}). 
Notice that,  as seen in (\ref{minimal}), the new scalar mode is a 
massless free field and it is possible to calculate the corresponding 
power spectrum  generated during
inflation, 
$P_{\nabla A}(k)=4\pi k^3\vert\nabla_\mu\A^{(s)\mu}_k\vert^2 $. In the
super-Hubble limit, we get  in a 
 quasi-de Sitter inflationary phase characterized by a slow-roll
parameter $\epsilon$:
\begin{eqnarray}
P_{\nabla A}(k)=\frac{9H_{k_0}^4}{16\pi^2}
\left(\frac{k}{k_0}\right)^{-4\epsilon}
\label{PE}
\end{eqnarray}
where $H_{k_0}$ is the Hubble parameter when the 
$k_0$ mode left the horizon \cite{EM1}. Notice that this
result implies that $\rho_A\sim (H_{k_0})^4$. The measured value of
the cosmological constant then requires $H_{k_0}\sim 10^{-3}$ eV,
which corresponds to an inflationary scale  $M_I\sim 1$ TeV.
Thus we see that the cosmological constant scale can be naturally
explained in terms of physics at the electroweak scale.
This is one of the most relevant aspects of the present model
in which, unlike existing dark energy theories based on scalar fields, 
dark energy can be generated without including any potential term
or dimensional constant. 

Since the field amplitude of the scalar state
remains frozen on super-Hubble scales, there is no modification
of Maxwell's equation on those scales. However, as the
amplitude starts decaying once the
mode enters the horizon in the radiation or matter eras, 
 the $\xi$-term in (\ref{EMeqexp}) generates an effective
current which can produce magnetic fields on cosmological
scales, as we will show in the following.

By passing, we notice that, in Minkowski spacetime, the  theory (\ref{actionF})
is completely equivalent to standard QED because, although
non-gauge invariant, the corresponding effective action is equivalent to the 
standard BRST invariant effective action of QED \cite{EM2}. This prevents from potential unobserved effects in accelerators because both theories lead to the same phenomenology in flat spacetime, being distinguishable only in curved backgrounds.

On the other hand, despite the fact that the homogeneous evolution in the present case
is the same as
in $\Lambda$CDM, the effective cosmological constant generated by the new scalar state fluctuates so that the evolution of metric perturbations could
be different. We have calculated the evolution of metric,
matter density and
electromagnetic perturbations \cite{EM3}. The propagation speeds
of scalar, vector and tensor perturbations are found
to be real and equal to the speed of light, so that the theory is
classically stable. Moreover, the three physical states carry positive energy so it is also quantum-mechanically stable.
On the other hand, it is
possible to see that all the post-Newtonian parameters
\cite{Will}
agree with those of General Relativity,  i.e. the theory is compatible
with all the local gravity constraints for any value
of the homogeneous background electromagnetic field at the same level of accuracy as General Relativity \cite{EM1,viability}. Concerning the evolution of
scalar perturbations, 
we find
that  the only relevant deviations with respect to $\Lambda$CDM
appear on large scales $k\sim H_0$ and that
they depend on the primordial
spectrum of electromagnetic fluctuations. However,
the effects on the CMB temperature and matter power spectra 
are compatible with observations except for very large primordial
fluctuations \cite{EM3}.

\section{Generation of cosmic magnetic fields}

It is interesting to note that the $\xi$-term 
can be seen, at the equations of motion
level, as a conserved current acting as a source of the usual
Maxwell field. To see this, we can write
$-\xi\nabla^\mu(\nabla_\nu A^\nu)\equiv J_{\nabla\cdot A}^\mu$
which, according to (\ref{minimal}), satisfies the conservation
equation $\nabla_\mu J_{\nabla\cdot A}^\mu=0$ and we can express
(\ref{EMeqexp}) as:
\begin{eqnarray}
\nabla_\nu F^{\mu\nu}=J^\mu_T
\end{eqnarray}
with $J^\mu_T=J^\mu+J^\mu_{\nabla\cdot A}$ and $\nabla_\mu
J^\mu_T=0$. Physically, this means that, while the new scalar mode
can only be excited gravitationally, once it is
produced it will generally behave as a source of electromagnetic
fields. Therefore,  the modified theory is described by 
ordinary Maxwell equations with an additional "external" current.
For an observer with four-velocity
$u^\mu$ moving with the cosmic plasma, 
it is possible to  decompose the Faraday tensor in its 
electric and magnetic parts  
as: $F_{\mu\nu}=2E_{[\mu}u_{\nu ]}+\frac{\epsilon_{\mu\nu\rho\sigma}}
{\sqrt{g}}B^\rho u^\sigma$, where $E^\mu=F^{\mu\nu} u_\nu$ and
 $B^\mu=\epsilon^{\mu\nu\rho\sigma} /(2\sqrt{g})F_{\rho\sigma}u_\nu$.
Due to the infinite conductivity of the plasma, 
Ohm's law $J^\mu-u^\mu u_\nu J^\nu=\sigma F^{\mu\nu} u_\nu$
implies  $E^\mu=0$.  
Therefore, in that case the only contribution  would come
from  the magnetic part.  
Thus, from Maxwell's equations, we get, for comoving observers in a FLRW metric
(see also \cite{FC}):
\begin{eqnarray}
\vec \omega\cdot \vec B=\rho_g^0
\label{mag}
\end{eqnarray}
where $\vec v=d\vec x/d\eta$ is the conformal
time fluid velocity, $\vec\omega=\vec \nabla\times \vec v$ is 
the fluid 
vorticity, $\rho_g^0=-\xi\partial_0(\nabla_\mu A^\mu)$ is the effective charge density today
whose power spectrum can be obtained from (\ref{PE}) (see \cite{BMminimal} for more details),  
and the $\vec B$ components scale as $B_i\propto 1/a$ as can be
easily obtained from $\epsilon^{\mu\nu\rho\sigma}F_{\rho\sigma;\nu}=0$
to the lowest order in $\vert\vec{v}\vert$. Thus,  the 
presence of the non-vanishing cosmic effective charge density necessarily 
creates both magnetic field and vorticity. 
Due to the 
presence of the 
effective current, we find that vorticity grows as $\vert\vec\omega\vert
\propto a$, from radiation era until present.

\begin{figure}[t]
\begin{minipage}{0.5\textwidth}
\includegraphics[width=8cm]{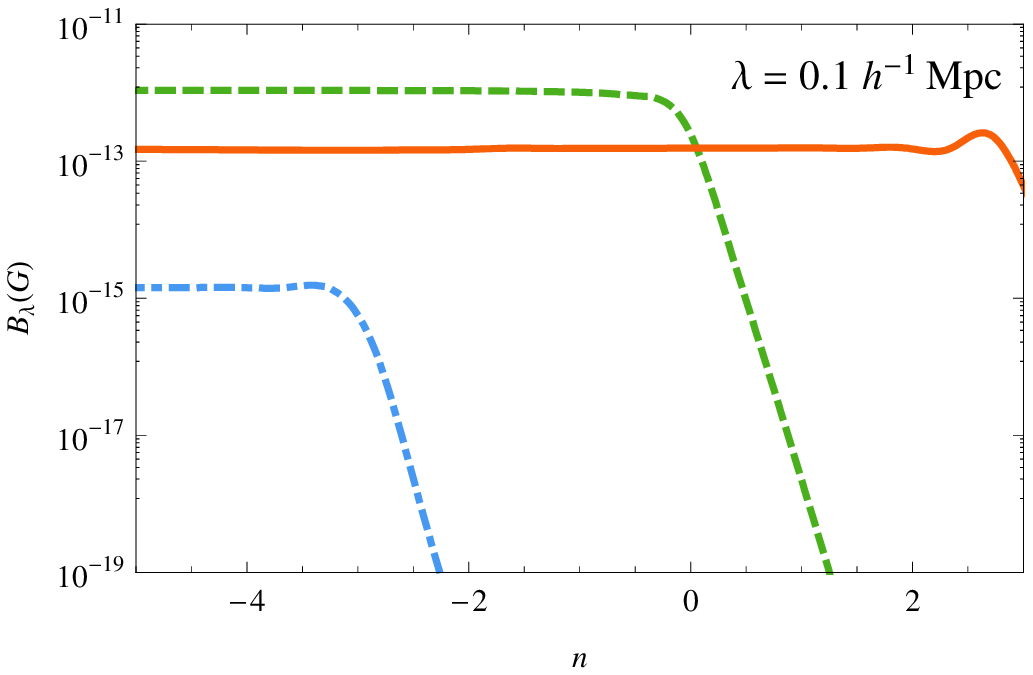}
\end{minipage}
\hfill
\begin{minipage}{0.5 \textwidth}
\includegraphics[width=8cm]{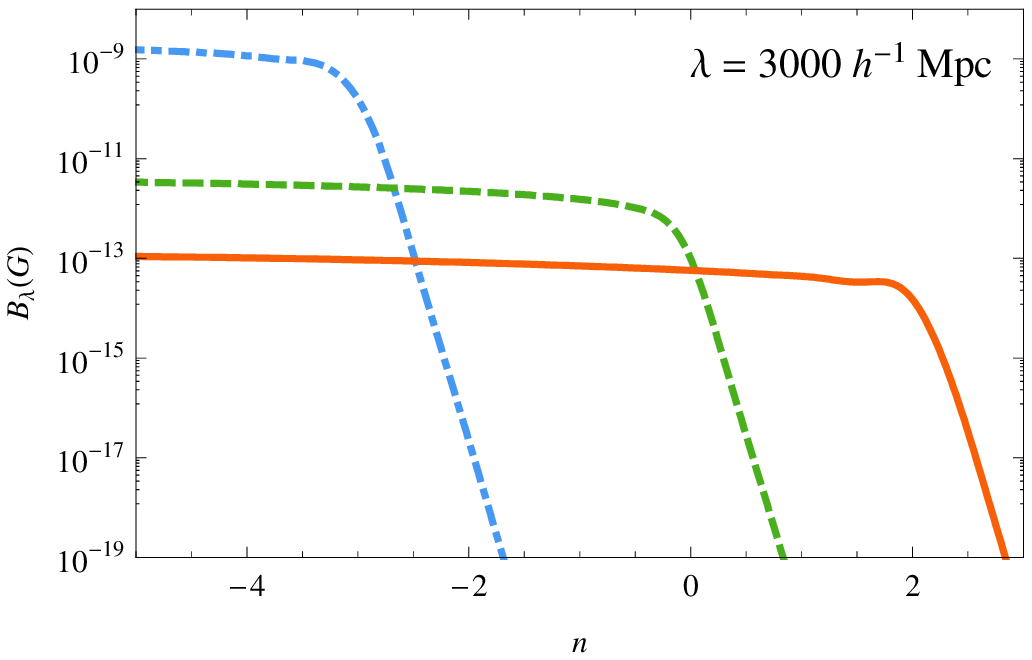}
\end{minipage} 
\caption{ Lower limits on the magnetic
fields generated  on galactic scales (left panel)
and Hubble horizon scales (right panel) in terms
of the magnetic spectral index $n$ for different values
of the vorticity spectral index $m$. Dot-dashed blue for $m=0$, dashed 
green
for $m\simeq -3$ and full red for $m\simeq -5$.}
\label{BMfig}
\end{figure}

 Using (\ref{mag}), it is possible to translate the existing
upper limits on vorticity coming from CMB anisotropies \cite{FC} into  
{\it lower} limits on the amplitude of the magnetic fields 
generated by this mechanism \cite{BMminimal}. In Fig. \ref{BMfig} we show the corresponding limits on the magnetic field at galactic and Hubble horizon scales for different spectral indices of magnetic field ($n$) and vorticity ($m$). We see that this mechanism allows to generate relatively strong magnetic fields on scales as large as the Hubble horizon and act as seeds for a galactic dynamo or even play the role of primordial fields and account for observations just by amplification due to the collapse and differential rotation of the protogalactic cloud. 

\section{Discussion}
We have discussed and extended electromagnetic theory in which we do not need to impose the Lorenz condition. In order to quantise the theory we introduce an additional scalar state which can be excited by gravitational fields. Indeed, fluctuations of such a state during an inflationary era at the electroweak scale generates an effective cosmological constant on super-Hubble scales with the correct value. This theory is free form both classical and quantum instabilities and is consistent with all local gravity tests at the same level as General Relativity, CMB and large scale structure observations. On the other hand, the sub-Hubble modes of the new state generated during inflation acts as an effective electromagnetic current so it can produce cosmological magnetic fields all the way to the Hubble horizon (but not beyond).  This allows to establish an important link between the problems of dark energy and cosmic magnetic fields. In fact, also non-minimal couplings have been considered in \cite{BMnonminimal} where a potential relationship between angular momentum and magnetic fields has been explored.

\vspace{0.1cm}
{\bf Acknowledgements}:
This work has been supported by MICINN (Spain) project numbers FIS 2008- 01323 and FPA 2008-00592, CAM/UCM 910309, MEC grant BES-2006-12059 and MICINN Consolider-Ingenio MULTIDARK CSD2009-00064. J.B. also wishes to thank support from the Norwegian Council under the YGGDRASIL project no 195761/V11.

\end{document}